\documentclass[11pt]{article}

\usepackage{geometry}
\usepackage{amsmath, amssymb, amsthm}

\usepackage{graphicx}

\usepackage{hyperref}

\usepackage{algorithm}
\usepackage{algorithmic}
\usepackage{bm}
\usepackage{todonotes}
\usepackage{hyperref}

\theoremstyle{plain} 

\theoremstyle{definition}

\theoremstyle{remark}
\newtheorem{remark}{Remark}

\usepackage{caption}

\begin{document}
\title{Consistent initialization of mixed-dimensional multiphysics models for fractured reservoirs under geomechanical constraints and field measurements}

\author{
  Jakub Wiktor Both\thanks{Center for Modeling of Coupled Subsurface Dynamics, Department of Mathematics, University of Bergen, Allégaten 41, 5007 Bergen, Norway. Email: \texttt{jakub.both@uib.no}.}
  \and
  Inga Berre\thanks{Center for Modeling of Coupled Subsurface Dynamics, Department of Mathematics, University of Bergen, Allégaten 41, 5007 Bergen, Norway. Email: \texttt{inga.berre@uib.no}.}
}

\date{} 
\maketitle

\abstract{Modeling coupled processes in fractured porous media -- flow, deformation, fracture mechanics, and thermal/chemical effects -- often relies on mixed dimensional multiphysics formulations. These systems are nonlinear and depend on physical states and state dependent material laws. While in-situ field measurements consistently describe the deformed equilibrium configuration, computational models typically start from an idealized reference configuration and require explicit initialization of the in-situ stress state. This mismatch complicates initialization and linearization of constitutive laws. As a consequence, due to the two scale nature of fractured media, this can induce large deviations in fracture aperture directly impacting flow predictions. To address this, a discrete fracture model is introduced whose constitutive laws are expressed with respect to the unknown equilibrium state. This is paired with a fixed point initialization strategy that consistently reconstructs the reference configuration, consistent with both geomechanical constraints and field measurements up to load-path dependence. This data-consistent strategy provides a foundation for extending models to more complex scenarios, including multiphase and multicomponent flow in fractured reservoirs.}

\section{Introduction}
Geophysical systems of societal and industrial relevance, such as geothermal reservoirs and fractured crystalline rock, exhibit a combination of nonlinear physics, heterogeneous material behavior, and intricate fracture networks. Their evolution is in general governed by tightly coupled thermo‑hydro‑mechanical-chemical (THMC) processes, while flow and transport pathways are strongly affected by the opening, closure, and shear behavior of individual fractures~\cite{vaezi2025review}. Accurate modeling of the associated non‑smooth mechanics, especially during transitions between open and contact states, pose therefore a substantial requirement for predictive modeling. Mixed‑dimensional discrete‑fracture–matrix (DFM) formulations have emerged as a consistent mathematical and computational framework to resolve these coupled phenomena while explicitly representing the geometry and physics of fracture networks~\cite{berre2019flow,stefansson2021fully}.

In subsurface applications, the fractured geometry and material properties obtained from field campaigns before injection and production operations correspond to an in‑situ equilibrium configuration shaped by the interplay of lithostatic loading, hydrostatic pressure, and regionally varying tectonic stresses~\cite{heidbach2018world}. However, aiming at initializing with the equilibrium state, most numerical geomechanics workflows begin from a stress‑free reference configuration and recover an equilibrated initial state through gravity or hydrostatic initialization~\cite{geosx_init_equil}. While this approach yields reasonable stress fields in the intact matrix, it can introduce significant inconsistencies in fractured domains: even small perturbations of fracture-normal stresses can substantially modify mechanical apertures and, consequently, fracture transmissivities, inconsistent with field measurements. As a result, initialization artifacts can overshadow genuine coupled THMC responses and severely distort flow predictions during forward simulations.

At the same time, multi-disciplinary reservoir characterization provides increasingly rich constraints. Borehole images, acoustic logs, well tests, and hydraulic measurements offer information on fracture orientation, roughness, fill, stress state, and effective transmissivity, while laboratory experiments deliver porosity, permeability, elastic moduli, and density~\cite{ma2021multi,finnila2021revisions}. This wealth of data enables reasonable estimates for reference values across various material parameters and constitutive relations, when linearized/expanded around the equilibrium state. However, coupled THMC models commonly linearize constitutive laws around a stress‑free configuration. This results in two important limitations:

\begin{enumerate}
    \item Initialization-induced aperture biases: equilibration routines may produce unphysically large fracture openings and artificially enhanced flow pathways.
    \item Limited integration of measurements: many parameters are challenging to incorporate, and particularly those associated with the in-situ state.
\end{enumerate}

In this work, we introduce a coupled poromechanics model for fractured porous media with frictional contact that incorporates reference states directly into all constitutive and kinematic definitions. This reference‑state‑aware formulation offers a systematic way to embed field measurements directly into both the choice of primary variables and the constitutive relations governing coupled processes. Together with the modeling approach, we propose a simple yet robust iterative initialization procedure that reconstructs and treats the unknown in‑situ equilibrium configuration as a part of the problem, consistent with geomechanical constraints loads and field measurements and taken as initial state for a forward model. With reference displacements holding the main complexity, this reduced perspective serves as boilerplate and can directly be applied to general THMC modeling.

The remainder of the article is organized as follows. Section~\ref{sec:model} presents the reference‑state‑aware mathematical model, including the choice of primary variables and constitutive structure. Section~\ref{sec:initialization} introduces the initialization strategy that reconstructs the in-situ equilibrium configuration from measurements and external forcing. Section~\ref{sec:example} provides numerical studies quantifying the impact of the modeling approach and equilibration strategy and discussion on path-dependence. Concluding remarks are given in Section~\ref{sec:remarks}.

\section{Poromechanics modeling around reference states\label{sec:model}}
Models for coupled multiphysics processes in geophysical fractured systems combine first‑principles governing equations (mass conservation, momentum balance in both matrix and fractures) with constitutive relations that capture the physics of flow, deformation, and fracture behavior (e.g., Darcy flow, poroelasticity, friction, and aperture–permeability relations). In this section, we build on these fundamental ingredients and introduce a formulation in which all constitutive laws explicitly incorporate a reference configuration.



Each physical quantity can, in principle, be defined relative to a reference configuration. The particular choice of primary variables and constitutive relations therefore utilizes a collection of reference states, which may be fully known, partially constrained by measurements, or entirely unknown. In this work, we assume for simplicity that the observed geophysical system is initially at rest, and we take this in‑situ equilibrium to define the reference configuration for all constitutive laws. Accordingly, initial/reference values are denoted with a subscript $0$. We employ the general modeling principle: \textit{Given the state $s$, each quantity of interest $q$ has its reference value $q_0$ at some reference state $s_0$ and is otherwise expressed as functional expansion around that reference state}
\begin{align}
\label{eq:modeling-principle}
q(s) =  Q_0(s-s_0) + q_0\cdot Q_1(s - s_0),
\end{align}
\textit{where $Q_0$ and $Q_1$ are potentially nonlinear and satisfy $Q_0(0)=0$ and $Q_1(0)=1$, enabling additive and multiplicative expansions.}


\subsection{Modeling assumptions}

The following assumptions will motivate the subsequent choices of constitutive laws: 1. The material is poroelastic and has a linear effective stress-strain relationship around the equilibrium state with the classical linear pore pressure following Biot's linear theory of poroelasticity. 2. Fluid flow is governed by linear Darcy flow across matrix and fractures with cubic-law fracture permeability. 3. In-situ porosity and fracture aperture are known and are varying linearly around the equilibrium state. 4. At initial time, the material is in equilibrium with geomechanical constraints imposed by gravitational forces, hydrostatic and lithostatic pressures and tectonic stresses.





\subsection{Mixed-dimensional geometry and computational domain\label{sec:domain}}

In this work, we consider a spatial configuration $\Omega\subset \mathbb{R}^d$, $d\in\{2,3\}$, (called matrix) as obtained from reservoir characterization. Fractures, denoted $\Gamma$, are represented as lower‑dimensional ellipsoidal surfaces embedded within $\Omega$; their possible intersections generate a hierarchy of progressively lower‑dimensional subdomains. Focusing directly on the impact of the reference states, we omit technical details of domains and operators for the transfer in between matrix and fractures (typically realized through mortar methodology) and for simplicity identify all lower-dimensional objects with the fracture. Details on mixed‑dimensional geometries can be found in \cite{berre2019flow,stefansson2021fully,keilegavlen2021porepy}. Overall, based on the modeling assumption of linear elasticity, this collection of domains also defines the reference computational domain, on which all partial differential equations and associated variables are defined.

\subsection{Primary variables and the special role of mechanical displacements\label{sec:primary-variables}}

To describe poromechanics, we adopt as primary variables the mechanical displacement of the matrix, $\bm{u}$, together with the fluid pressures in the matrix and fractures, $p$ and $p_f$. Mechanical deformation is governed by the total stress, $\bm{\sigma}$, within the matrix complying with tractions, $\bm{\lambda}$, acting on fracture interfaces. Fluid flow on the other hand is controlled by fluxes within the matrix, fractures, and across the matrix-fracture interfaces.

\begin{figure}[!ht]
    \centering
    \includegraphics[width=0.9\linewidth]{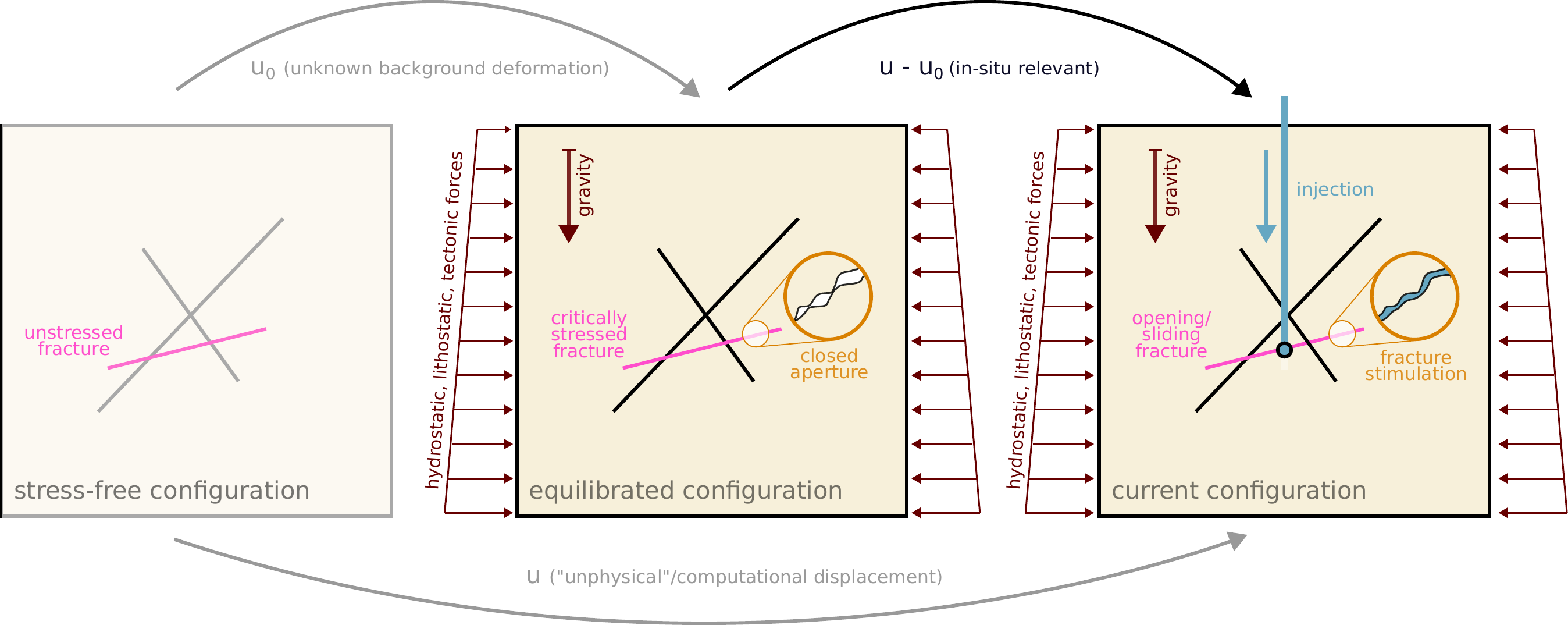}
    \caption{Illustration of spatial stress-free, equilibrium and dynamic configurations with associated notion of mechanical displacement in between these.}
    \label{fig:three-configurations}
\end{figure}

Mechanical displacements play a special role because they are in general defined as being relative to a specific spatial reference configuration. We distinguish three configurations, cf. Figure~\ref{fig:three-configurations}: the stress‑free computational reference configuration (cf.~\ref{sec:domain}), the equilibrium configuration complying with geomechanical constraints (hydrostatic/lithostatic pressures and tectonic horizontal stresses), and the current configuration, which evolves in time due to external forcing such as fluid injection. With variables associated to the computational reference configuration, we denote the displacement between the stress-free and equilibrium configuration by $\bm{u}_0$, and $\bm{u}$ is associated with the current configuration. Consequently, $\bm{u} - \bm{u}_0$ denotes in-situ measurable displacement and is of main relevance in the field settings. Despite the linearizing assumptions of linear elasticity with infinitesimal strains, implying that the three configurations are practically identical, this distinction remains essential for consistently relating displacements to stresses, fracture aperture evolution, and the resulting modifications of effective hydraulic properties.

The remaining variables have an absolute character. The introduction of a reference pressure $p_0$, reference stresses $\bm{\sigma}_0$ and $\bm{\lambda}_0$, and reference fluxes (zero assuming geomechanical constraints that do not induce background flow) --  associated with the equilibrated reference state -- is conceptually simpler and results in an additive split. The direct access to the deviation from the equilibrium state as $p-p_0$ etc. will be central in consistent constitutive modeling and integration of field measurements.

\subsection{Constitutive modeling of poroelastic materials\label{sec:constitutive-laws}}

We consider the reference state $s_0 = (\bm{u}_0, p_0, \bm{\sigma}_0, \bm{\lambda}_0)$ and adapt central constitutive relations for the modeling of poromechanic in fractured media, complying with~\eqref{eq:modeling-principle}. This enables meeting the co-existing objectives of linearizing parameter laws around reference states and fitting laws to field measurements.

\vspace{0.1cm}

\begin{figure}[ht!]
    \centering
    \includegraphics[width=0.9\linewidth]{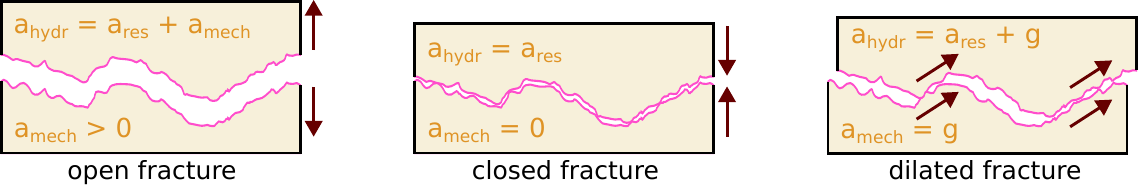}
    \caption{Fracture aperture modeling: Open, closed, and shear-dilated fracture modes. In closed state, positive residual aperture $a_\mathrm{res}$ remains available. In open and dilated states, a non-negative mechanical aperture contributes, while in the latter, it is governed through the fracture gap model, cf.~\eqref{eq:contact-normal-1}--\eqref{eq:contact-normal-3}.}
    \label{fig:aperture}
\end{figure}

\noindent
\textbf{Constitutive modeling of fracture.} The fracture aperture highlights the multi-scale character of fractured media and has a central impact on hydraulic properties through modeling of fracture permeability~\cite{berre2019flow}. We distinguish between mechanical and hydraulic apertures, resulting from the fact that despite in closed state, roughness of fractures retains residual volume for fluid to flow and thus contributes to hydraulic properties; the nuances between different notions of apertures is illustrated in Fig.~\ref{fig:aperture}. 
Overall, we consider:


\begin{subequations}\label{eq:constitutive-laws-aperture}
    
\begin{itemize}

\item \underline{Mechanical aperture} (non-negative by construction and complementarity):
\begin{align}
\label{eq:mechanical-aperture}
    a_\mathrm{mech} &= a_\mathrm{mech,0} + [\bm u - \bm u_0]_n,
\end{align}
where $[\cdot]_n$ denotes the jump across a fracture in normal direction. \vspace{2mm}

\item \underline{Fracture gap} (aperture increase, e.g., due to deformation history in terms of shear dilation~\cite{willis1996progress}, extandable to first-order compressibility effects~\cite{barton1985strength} and others):
\begin{align}
\label{eq:fracture-gap}
    g &= g_0 
    + \mathrm{tan}(\theta_\mathrm{dil})\cdot \big|[\bm u - \bm{u}_0]_t\big| 
\end{align}
where $[\cdot]_t$ is the tangential jump, and $g_0$ identifies an initial gap. \vspace{2mm}

\item \underline{Hydraulic aperture} (effectively available pore space in fracture):
\begin{align}
    a_{\mathrm{hydr}} &= a_\mathrm{res} + a_\mathrm{mech} 
\end{align}

\item \underline{Fracture permeability} (correlated with the aperture, here via a cubic-law~\cite{berre2019flow}):
\begin{align}
    \kappa_f &= a_{\mathrm{hydr}}^2 / 12.
\end{align}


\end{itemize}

\end{subequations}

\noindent
\textbf{Constitutive modeling of matrix properties.} Hydraulic and mechanical properties in relation to associated reference states include:
\begin{subequations}\label{eq:constitutive-laws}
    
\begin{itemize}

\item \underline{Total stress} (based on classical effective stress theory~\cite{coussy2004poromechanics}):
\begin{align}
\label{eq:sigma}
    \bm{\sigma} &= \bm\sigma_0 + \mathbb{C}\bm{\varepsilon}(\bm{u} - \bm{u}_0) - \alpha (p - p_0) \bm{I},
\end{align}
where it is important to note that the reference stress $\bm{\sigma}_0$ is not known a priori, and is characterized by geomechanical constraints. Although allowing for distinct modeling, simplifying the presentation, we employ a linear stress-strain relation for the deformation of the stress-free to the equilibrium spatial configuration and set
\begin{align}
\label{eq:background-stress}
\bm\sigma_0 = \mathbb{C}\bm{\varepsilon}(\bm{u}_0) - \alpha p_0 \bm{I}.
\end{align}

\item \underline{Porosity} (employing standard linearized poroelasticity~\cite{coussy2004poromechanics}):
\begin{align}
    \phi &= \phi_0 + \alpha \nabla \cdot (\bm{u} - \bm{u}_0) + \frac{1}{M}(p - p_0),
\end{align}
where $\phi_0$ typically would be heterogeneous and depending on the heterogeneous background stress $\bm{\sigma}_0$ and background pressure $p_0$, e.g., following again standard theory and based on some (stress-free) ex-situ measure porosity $\phi_\mathrm{ref}$ 
\begin{align}
\phi_0 = \phi_\mathrm{ref} + \frac{\alpha^2}{K_\mathrm{dr}} \mathrm{tr}(\bm{\sigma}_0) + \frac{1}{M}p_0
\end{align}

\item \underline{Fluid density} (employing a simple thermodynamical law):
\begin{align}
    \rho & = \rho_\mathrm{ref} \cdot \mathrm{exp}(c(p-p_\mathrm{ref})),
\end{align}
where $p_\mathrm{ref}$ is a reference pressure associated with the reference density $\rho_\mathrm{ref}$, typically obtained during ex-situ laboratory testing and independent of $p_0$. 
\end{itemize}

\end{subequations}

\subsection{Coupled model characterizing reference and current states\label{sec:full-model}}


The coupled model can be split into the subproblems of fluid flow, mechanical deformation, and contact mechanics, while we implicitly will employ the reference-aware constitutive laws introduced in Section~\ref{sec:constitutive-laws}. Fluid flow is modeled using a DFM modeling principle combining conservation of mass, Darcy's law, and direct impact of hydraulic properties such as matrix porosity, hydraulic aperture, and fracture permeability; for more details, we refer to~\cite{berre2019flow,stefansson2021fully}. We employ the mixed-dimensional model
\begin{subequations}
\label{eq:pde}
\begin{alignat}{5}
\label{eq:mass-conservation:matrix}
\partial_t\!\big(\rho\phi\big)
+ \nabla \cdot \left(-\frac{\rho\kappa}{\mu_f}(\nabla p - \rho \mathbf{g})\right) 
&= h
&\quad &&\text{on } \Omega &\ &\text{(domain)}\\[0.25em]
\label{eq:mass-conservation:fracture}
\partial_t\!\big(\rho\,a_{\mathrm{hydr}}\big)
+ \nabla \cdot \left(-\frac{\rho\,a_{\mathrm{hydr}}\,\kappa_f}{\mu_f}(\nabla p_f - \rho \mathbf{g})\right)
&= h_f +  q_\perp
&\quad &&\text{on } \Gamma &&\text{(domain)}\\[0.25em]
\label{eq:mass-conservation:interface:flux}
-\frac{\rho\kappa}{\mu_f}\left(\nabla p - \rho \mathbf{g}\right)\!\cdot\! \mathbf{n}
&= q_{\perp}
&\quad &&\text{on } \Gamma &&\text{(boundary)}\\[0.25em]
\label{eq:mass-conservation:interface:law}
-\dfrac{2\kappa_{\perp}}{a_{\mathrm{hydr}}}\,(p - p_f) 
&= q_{\perp}
&\quad &&\text{on } \Gamma &&\text{(interface)}.
\end{alignat}
The fracture $\Gamma$ serves multiple purposes (indicated in parentheses), and thus  $q_\perp$ acts both as boundary condition and source. Here, we employ a constant matrix permeability $\kappa$, normal permeability $\kappa_\perp$, fluid viscosity $\mu_f$, external source terms $h$ and $h_f$.

Mechanical deformation is governed by the balance of linear momentum combined with the Biot theory combined with contact mechanics introducing a force balance
\begin{alignat}{3}
\label{eq:linear-momentum:matrix}
    -\nabla \cdot \bm{\sigma} &= (\phi\rho + (1-\phi)\rho_s)\bm{g} &\quad&\text{on } \Omega\\
\label{eq:linear-momentum:interface}
    \bm\sigma \bm n &= \bm{\lambda} - p_f\bm n &&\text{on }\Gamma,
\end{alignat}
where $\rho_s$ denotes the constant solid density. Contact mechanics modeling splits into normal compliance conditions (non-penetration wrt.\ the fracture gap) on $\Gamma$
\begin{align}
    \label{eq:contact-normal-1}
    a_\mathrm{mech} - g &\geq 0\\
    \label{eq:contact-normal-2}
    \lambda_{n} &\leq 0\\
    \label{eq:contact-normal-3}
    \lambda_{n} (a_\mathrm{mech} - g) &= 0.
\end{align}
and friction modeling (here Coulomb's law) in tangential direction (also on $\Gamma$)
\begin{align}
    |\bm{\lambda}_{t}| + \mu \lambda_{n} &\leq 0 \\
    [\bm \dot{u}]_t &= 0 \qquad \text{if } |\bm{\lambda}_{t}| < -\mu \lambda_{n}\\
    [\bm \dot{u}]_t &= \xi \bm{\lambda}_{t} \qquad \xi \geq 0 \text{ otherwise.}
\end{align}
To ensure consistency with geomechanical constraint, the governing equations are paired with boundary conditions for specified $\partial\Omega_\mathrm{D,M} \cup \partial\Omega_\mathrm{N,M} = \partial\Omega_\mathrm{D,F} \cup \partial\Omega_\mathrm{N,F} = \partial\Omega$ with outer normal $\bm{n}$ and in-situ measured data $\bm{\sigma}_\mathrm{lith/tect}$ and $p_\mathrm{hydr}$ 
\begin{alignat}{3}
\bm{u} &= \bm{u}_0 &\quad&\text{ on }\partial\Omega_\mathrm{D,m} \label{eq:bc-u}\\
\bm{\sigma}\bm{n} &= \bm{\sigma}_\mathrm{lith/tect}\bm{n} &&\text{ on }\partial\Omega_\mathrm{N,m} \label{eq:bc-sigma}\\
p &= p_\mathrm{hydr} &&\text{ on }\partial\Omega_\mathrm{D,f} \\
\nabla p \cdot \bm{n}  &= 0 &&\text{ on }\partial\Omega_\mathrm{N,f}
\end{alignat}
as well as initial conditions, assumed to be identical with the reference state,
\begin{alignat}{3}
    \bm{u} &= \bm{u}_0 &\quad &\text{on }\Omega\\
    p &= p_0 &\quad &\text{on }\Omega.
\end{alignat}
\end{subequations}

\begin{remark}[Quasi-static evolution\label{remark:quasi-static}] On geological time scales, reservoir deformation proceeds through a sequence of near-static states, with the background stress and deformation adjusting between these stages. This quasi-static character is represented in model~\eqref{eq:pde}, which is formulated relative to a fixed reference state in time. In view of load-path dependence, cf.\ Remark~\ref{remark:nonunique-u0}, the background displacement $\bm{u}_0$ is not uniquely determined from first principles, and the absence of detailed geological history prevent a unique identification.
\end{remark}

\begin{remark}[Non-unique characterization of the reference state\label{remark:nonunique-u0}] 
The above model~\eqref{eq:pde} implicitly assumes consistency between the reference state and the governing equations at initial time. Thus, at initial time the contact conditions reduce to
\begin{align*}
    \lambda_{0,n} \leq 0\quad \text{and} \quad |\bm{\lambda}_{0,t}| + \mu\lambda_{0,n} \leq 0 \qquad\text{on }\Gamma.
\end{align*}
Since no restriction is imposed on the stationary displacement $\bm{u}_0$ itself, and a complete variational structure is lacking, the resulting equilibrated background state is generally non-unique. This feature is consistent with both the auxiliary role played by the computational reference configuration, cf.\ Section~\ref{sec:primary-variables}, and the hysteretic behavior of systems governed by Coulomb friction resulting in non-uniqueness~\cite{klarbring1990examples}. 
\end{remark}

\begin{remark}[Consistency of the modeling approach]
The governing equations in~\eqref{eq:pde} are written in terms of absolute stresses, forces, and fluxes, and do not propagate reference states through the balance laws. Instead, the reference configuration enters only through the constitutive relations, which ensures that field data and initial conditions can be incorporated in a consistent manner. At the initial time, the system is at rest, implying $\bm{\sigma} = \bm{\sigma}_0$, $\bm{\lambda} = \bm{\lambda}0$, and a stationary flow field with $p_0$ equal to the hydrostatic pressure (up to discretization effects). Likewise, the background stress obtained from the poroelastic model~\eqref{eq:background-stress} reflects the boundary conditions applied to $\partial\Omega$ while respecting the fracture geometry. This construction guarantees that any available measurements of $a{\mathrm{mech},0}$ or $\phi_0$ can be imposed coherently, including in regions that are effectively traction–free. In particular, open fractures may be assigned any admissible reference aperture consistent with the contact conditions at initial time.
\end{remark}

\begin{remark}[Numerical discretization]
We consider a cell-centered finite volume method tailored towards mixed-dimensional DFM modeling conforming to fracture networks, combined with an implicit Euler discretization; the details can be reviewed in~\cite{stefansson2021fully}. The numerical methodology is implemented in PorePy~\cite{keilegavlen2021porepy,stefansson2024flexible}. At each discrete time $t$, the system to be solved reads
\begin{align}\label{eq:time-stepping}
\mathbf{G}(\mathbf{X};\ \mathbf{X}_\mathrm{prev},\ \mathbf{X}_0,\ t) = \mathbf{0},
\end{align}
where $\mathbf{X}$, $\mathbf{X}_\mathrm{prev}$, and $\mathbf{X}_0$ denote the current, previous and reference states. 

\end{remark}


\section{Effective initialization strategies for obtaining reference states\label{sec:initialization}}

To use the formalism introduced in Section~\ref{sec:model}, the (discretized) stationary state is consistently characterized by the condition
\begin{align}
\label{eq:full-characterization-X_0}
\mathbf{G}(\mathbf{X}_0;\ \mathbf{X}_0,\  \mathbf{X}_0,\ t_0) = \mathbf{0}.
\end{align}
To effectively compute a consistent initial state $\mathbf{X}_0$, we employ a simple iterative initialization procedure that can be interpreted as a pseudo–time-stepping scheme consistent with~\eqref{eq:time-stepping} and reflecting the quasi-static character of geological systems, cf.\ Remark~\ref{remark:quasi-static}. A key advantage of this strategy is that it reuses the original model $\mathbf{G}$ without modification. The iteration merely updates $\mathbf{X}_0$ while retaining the same primary variables as in $\mathbf{G}$. In this way, we avoid treating $\mathbf{X}$ and $\mathbf{X}_0$ as simultaneous unknowns in~\eqref{eq:full-characterization-X_0}, circumvent the inherent non-uniqueness of the reference state, cf.\ Remark~\ref{remark:nonunique-u0}, and ensuring a consistent initialization.
\begin{algorithm}
\caption{Iterative Solution for finding $X_0$\label{alg:initialization-I}}
\begin{algorithmic}[1]
\STATE Set $\mathbf{X}_0^{(0)} = 0$ and select a tolerance $\varepsilon$
\FOR{$n = 1, 2, \ldots$}
    \STATE Solve $\mathbf{G}(\mathbf{X}_0^{(n)}; \mathbf{X}_0^{(n-1)}, \mathbf{X}_0^{(n-1)}) = 0$ for $X_0^{(n)}$ (e.g. using a semi-smooth Newton method)
    \IF{$|X_0^{(n)} - X_0^{(n-1)}| < \varepsilon$}
        \STATE \textbf{break}
    \ENDIF
\ENDFOR
\end{algorithmic}
\end{algorithm}



\begin{remark}[Improved initial guesses\label{remark:improved-initial-guess}]
Instead of employing trivial initial guesses for $\mathbf{X}_0^{(0)}$, cf.\ Alg.~\ref{alg:initialization-I}, improved guesses, e.g., analytical expressions of the hydrostatic pressure combined with the resulting displacement decoupled from the flow problem, are expected to improve the numerical efficiency.
\end{remark}

\begin{remark}[Alternative models\label{remark:alternative-models}]
To improve convergence during initialization, one may employ simplified variants of the model $\mathbf{G}$, e.g., in cases when several constitutive relations reduce to constants and the equilibrium configuration is (quasi-)stationary (assuming no background flow and simple background deformation), which typically lowers the degree of nonlinearity. Likewise, ideas from homotopy continuation could be used to transition gradually between models, for instance through staged application of external loads or constitutive parameters. 
\end{remark}

\section{Numerical consistency check in light of path-dependence \label{sec:example}}

\begin{figure}
    \centering
    \includegraphics[width=0.9\textwidth]{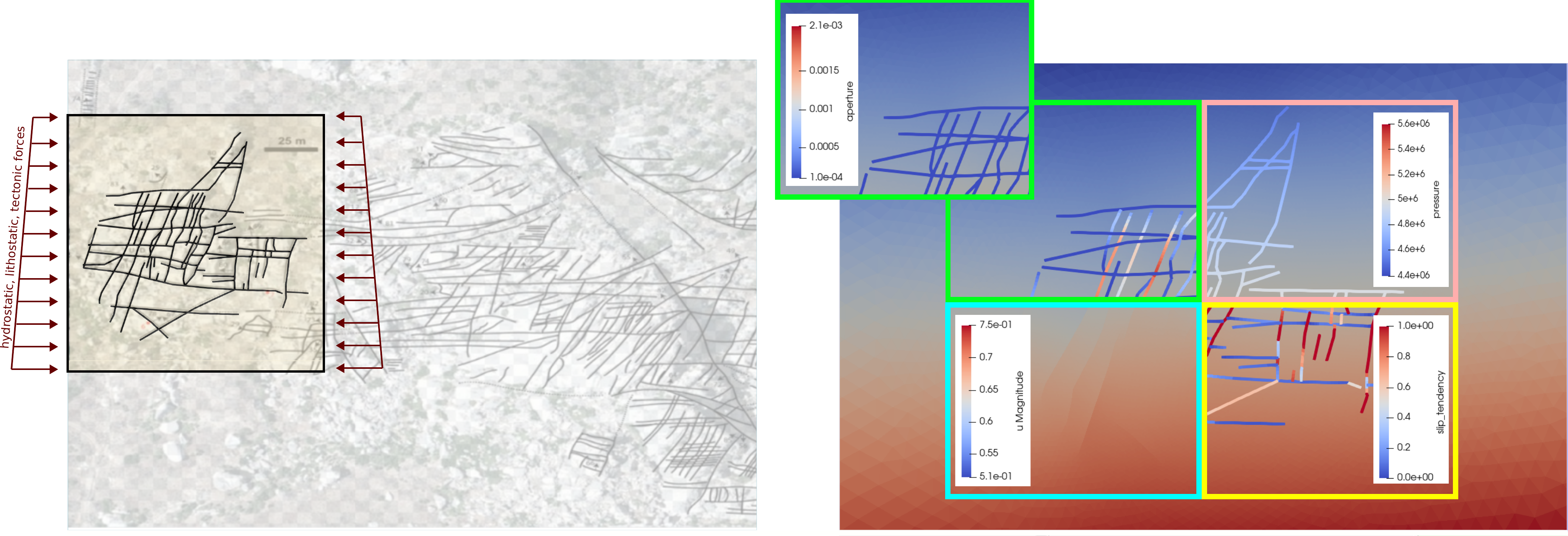}
    \caption{\textit{Left:} Salt Cove, modified excerpt from~\cite{finnila2021revisions}, the box marks the considered ca. 150 $\times$ 150 m sized fracture network embedded in a vertical 2d domains and subject to geomechanical constraints (hydrostatic, lithostatic and horizontal stress for 3000 - 4000 m) \textit{Right:} Equilibrated state with highlighted boxes; consistent and inconsistent apertures (green top left), background deformation (blue, bottom left, and outside the boxes), fracture pressure (pink, top right), slip tendency (yellow, bottom right).
    \label{fig:salt-cove}}
\end{figure}

We consider a two-dimensional test case based on a fracture network extracted from an outcrop at Salt Cove near the Utah FORGE site~\cite{finnila2021revisions}; see Fig.~\ref{fig:salt-cove}. The outcrop is interpreted as a vertical cross section embedded in a 1000 m $\times$ 1000 m domain. Lithostatic and hydrostatic pressures increasing with depth, corresponding to conditions of 3000–4000 m, are applied along the model boundaries. Horizontal stresses are prescribed as 25\% of the lithostatic pressure. Material parameters follow~\cite{Both2025EGC}. Specifically, we employ $a_\mathrm{mech,0}=0$ m, $\theta_\mathrm{dil}=0.1$, $\alpha=1$, $\kappa=4.35e-6$ D, $\kappa_\perp=4.35e-6$ D, $a_\mathrm{res} = 1e-4$ m, $\phi_0 \equiv 1.36e-2$, $G=16.8e9$ Pa and $\lambda=19.73e9$ Pa (defining $\mathbb{C}$), $\rho_s=2653$ kg/m$^3$, $\mu=0.6$ and properties of water for the fluid. A pseudo time step of 1 year is employed.

\paragraph{Consistent modeling of the reference state.} We begin by examining the consistent construction of the reference state using the initialization strategy described in Alg.~\ref{alg:initialization-I}. To enhance convergence, we initialize the pressure field with an analytical hydrostatic profile and, in the first iteration, solve a mechanics-contact problem subjected to this fluid pressure but decoupled from the flow equation; see Remark~\ref{remark:improved-initial-guess} and Remark~\ref{remark:alternative-models}. The final equilibrium configuration (after 5 iterations) is visualized in Fig.~\ref{fig:salt-cove}. 

\begin{figure}
    \centering
    \includegraphics[width=0.9\textwidth]{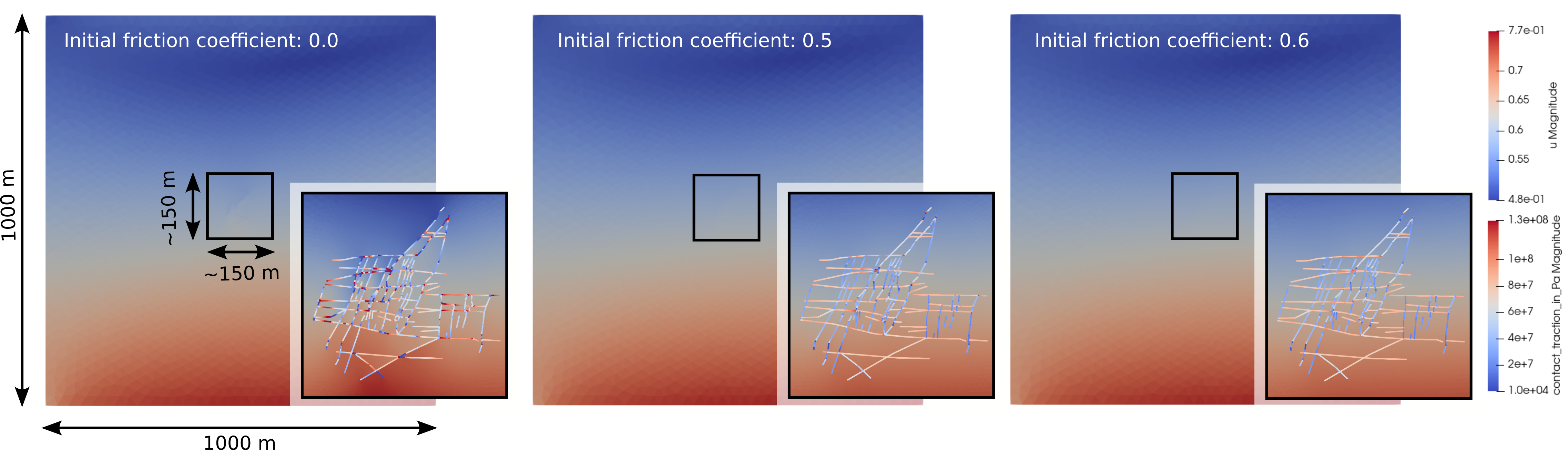}
    \caption{Path dependence of fractured systems under quasi-static friction. Final states after initialization with different initial friction values: 0.0 \textit{(left)}, 0.5 \textit{(center)}, 0.6 \textit{(right)}. Consistency in terms of far-field deformation, near-field deformation and contact traction shows (significant) differences.
    \label{fig:G0}}
\end{figure}

\paragraph{Path-dependence.} To illustrate the influence of load-path dependence, and thereby the inherent difficulty, yet importance, of reconstructing geological history in the absence of direct historical data, we consider a simplified scenario based on Remark~\ref{remark:alternative-models}. Motivated by fracture healing and the associated increase in frictional strength over time~\cite{dieterich1972time}, we prescribe a friction coefficient that varies over the pseudo–time steps: the system is initialized with reduced friction values, which are then restored to the target friction after three iterations. The resulting deformation field is shown in Fig.~\ref{fig:G0}. While the final states are (almost) indistinguishable in the far-field and remain consistent with geomechanical constraints, they exhibit significant differences in the near-fracture region. This behavior aligns with the inherent ambiguity in reconstructing the background deformation field, as discussed in Remark~\ref{remark:nonunique-u0}.

\section{Concluding remarks\label{sec:remarks}}

Field‑scale simulations are essential for planning and monitoring geothermal developments, and the increasing availability of reservoir characterization data, ranging from material parameters to detailed geomechanical constraints, makes it crucial to incorporate these measurements consistently into numerical models. In this work, we proposed a systematic modeling framework based on constructing reference states through equilibrium configurations, combined with an effective equilibration strategy. A central feature of this approach is that the same computational model used for forward simulations can also be used to generate the initial equilibrium state, thereby ensuring consistency with both field measurements and underlying geomechanical principles. 

While the focus here has been on poromechanics in fractured media, the main conceptual challenge lies in the treatment of displacement as a primary variable. As a result, extending the methodology to broader THMC models follows naturally from the same modeling principles, although multiphase–multicomponent settings would require coupling to more elaborate thermodynamic descriptions. The use of reference states and the additive decomposition in~\eqref{eq:modeling-principle} also raises the question of whether formulations based on deviations of primary variables can improve the nonlinear robustness of mixed‑dimensional THMC models, where convergence difficulties are well known~\cite{nevland2025augmented}. Looking forward, several aspects deserve further investigation. A more complete characterization of complex reservoirs and improved data integration workflows will be essential for reducing uncertainty in estimates of near‑fracture stress states, underscoring the value of borehole‑scale stress analyses, stress observations~\cite{heidbach2018world}, and hybrid/data‑integrated modeling approaches~\cite{gaucher2015induced}. Finally, the intrinsic ambiguity associated with the reference deformation state, as highlighted in this work, calls for deeper analysis to determine its consequences for model calibration, uncertainty quantification, and risk evaluation. Overall, the proposed reference‑state–based framework provides a consistent and extensible foundation for data‑integrated simulations of fractured geothermal reservoirs, and it offers a promising pathway toward more robust and physically grounded THMC modeling strategies.

\paragraph{Acknowledgment.}
This project has received funding from the European Research Council (ERC) under the
European Union’s Horizon 2020 research and innovation program (grant agreement No.
101002507). The authors thank Eirik Keilegavlen and Ivar Stefansson for their feedback.

%

\paragraph{Data availability statement} All data and materials to reproduce the numerical experiments are openly available at Zenodo \href{https://doi.org/10.5281/zenodo.18833363}{https://doi.org/10.5281/zenodo.18833363}


\bibliographystyle{plain}
\bibliography{references}
\end{document}